\documentclass[preprint,showpacs,titlepage,aps,prd,
tightenlines,amsmath,byrevtex,nofootinbib]{revtex4}

\usepackage{graphicx}

\def\lsim{\raise0.3ex\hbox{$\;<$\kern-0.75em\raise-1.1ex
\hbox{$\sim\;$}}}
\def\gsim{\raise0.3ex\hbox{$\;>$\kern-0.75em\raise-1.1ex
\hbox{$\sim\;$}}}

\begin{document}

\preprint{hep-ph/0607284}
\preprint{Fermilab-Pub-06-231-T}
%%%%%%%%%%%%%%%%%%%%%%%%%%%%%%%%%%%%%%%%%%%%%%%%%%%%%%%%%%%%%%%%%%%%%%%%%
\title{Determining Neutrino Mass Hierarchy by Precision Measurements in 
Electron and Muon Neutrino Disappearance Experiments} 
%%%%%%%%%%%%%%%%%%%%%%%%%%%%%%%%%%%%%%%%%%%%%%%%%%%%%%%%%%%%%%%%%%%%%%%%% 

\author{H.~Minakata$^{1,2}$}
%\author{Hisakazu Minakata$^{1,2}$}
\email{minakata@phys.metro-u.ac.jp}
\author{H.~Nunokawa$^{2}$}
%\author{Hiroshi Nunokawa$^{2}$}
\email{nunokawa@fis.puc-rio.br} 
\author{S.~J.~Parke$^{3}$}
%\author{Stephen Parke$^{3}$}
\email{parke@fnal.gov} 
\author{R.~Zukanovich Funchal$^{4}$}
%\author{Renata Zukanovich Funchal$^{4}$}
\email{zukanov@if.usp.br}
\affiliation{
$^1$Department of Physics, Tokyo Metropolitan University \\ 
1-1 Minami-Osawa, Hachioji, Tokyo 192-0397, Japan \\
$^2$Departamento de F\'{\i}sica, Pontif{\'\i}cia Universidade Cat{\'o}lica 
do Rio de Janeiro, C. P. 38071, 22452-970, Rio de Janeiro, Brazil \\
$^3$Theoretical Physics Department,
Fermi National Accelerator Laboratory, 
P.\ O.\ Box 500, Batavia, IL 60510, USA \\
$^4$Instituto de F\'{\i}sica, Universidade de S\~ao Paulo, 
 C.\ P.\ 66.318, 05315-970 S\~ao Paulo, Brazil
}

\date{\today}

\vglue 1.6cm
%%%%%%%%%%%%%%%%%%%%%%%%%%%%%%%%%%%%%%%%%%%%%%%%%%%%%%%%%%%%%%%%%%%%
%    Abstract
%%%%%%%%%%%%%%%%%%%%%%%%%%%%%%%%%%%%%%%%%%%%%%%%%%%%%%%%%%%%%%%%%%%%
\begin{abstract}
  Recently a new method for determining the neutrino mass hierarchy by
  comparing the effective values of the atmospheric $\Delta m^2$
  measured in the electron neutrino disappearance channel, $\Delta
  m^2(\text{ee})$, with the one measured in the muon neutrino
  disappearance channel, $\Delta m^2(\mu \mu)$, was proposed.  If
  $\Delta m^2(\text{ee})$ is larger (smaller) than $\Delta m^2(\mu
  \mu)$ the hierarchy is of the normal (inverted) type.  We re-examine
  this proposition in the light of two very high precision
  measurements: %(a few per mil):
  $\Delta m^2(\mu \mu)$ that %can
  may be accomplished by the phase II of the Tokai-to-Kamioka (T2K)
  experiment, for example, and $\Delta m^2(\text{ee})$ that can be
  envisaged using the novel M\"ossbauer enhanced resonant
  $\bar{\nu}_{e}$ absorption technique.  Under optimistic assumptions
  for the systematic uncertainties of both measurements, we estimate
  the parameter region of $(\theta_{13}, \delta)$ in which the mass
  hierarchy can be determined.  If $\theta_{13}$ is relatively large,
  $\sin^2 2\theta_{13} \gsim 0.05$, and both of $\Delta
  m^2(\text{ee})$ and $\Delta m^2(\mu \mu)$ can be measured with the
  precision of $\sim 0.5$ \% it is possible to determine the neutrino
  mass hierarchy at $>$ 95\% CL for $0.3 \, \pi \lsim \delta \lsim 1.7 \,
  \pi$ for the current best fit values of all the other oscillation
  parameters.
%  For slightly different values of these parameters, currently allowed by 
%  data, the sensitivity can  be even higher.
\end{abstract}

%\pacs{14.60.Pq,25.30.Pt,76.80.+y}

\maketitle

%%%%%%%%%%%%%%%%%%%%%%%%%%%%%%%%%%%%%%%%%%%%%%%%%%%%%%%%%%%%%%%%%%%%%%%%%%%
%\section{Introduction}
%\label{introduction}
%%%%%%%%%%%%%%%%%%%%%%%%%%%%%%%%%%%%%%%%%%%%%%%%%%%%%%%%%%%%%%%%%%%%%%%%%%%

In spite of the great progress that has been made in recent years 
there still remains a number of important questions regarding 
the nature of neutrinos.  
One such unknown is whether the neutrinos have a 
normal or inverted mass hierarchy.
%Among them, the problem of
%the neutrino mass hierarchy is one of the most important.  
This unresolved ambiguity in the neutrino mass pattern is usually
phrased as follows: if the neutrino masses are labeled 
as $m_1$ ($m_3$) for
the mass of the neutrino state with the greatest (least) $\nu_e$
component then for the normal hierarchy $m_3 > m_2 > m_1$ and 
for the inverted hierarchy $m_2 > m_1 > m_3$.
The determination of this pattern must shed light on the secret 
of how the lepton sector 
is organized and may testify to the underlying symmetries by which the 
structure of neutrino masses and lepton flavor mixing are prescribed 
\cite{review}. 

It is widely recognized that it is difficult to determine the neutrino
mass hierarchy by any of the ongoing and the near future neutrino
oscillation experiments including the atmospheric neutrino observation
by Super-Kamiokande \cite{SKI_atm}, MINOS~\cite{MINOS},
OPERA~\cite{OPERA}, and T2K (Tokai-to-Kamioka)~\cite{JPARC}. 
The most studied method for
determining the mass hierarchy within the scope of the future neutrino
experiments is to explore the earth matter effect in long-baseline
accelerator experiments~\cite{LBL_matter}.  
Since the way the
matter effect interferes with vacuum oscillation depends
upon the sign of $\Delta m^2_{31} \equiv m^2_3 - m^2_1$, the
degeneracy between the normal and the inverted mass hierarchies can be
lifted in matter.
The NO$\nu$A~\cite{NOVA} and the 
T2KK (Tokai-to-Kamioka-Korea)~\cite{T2KK} experiments 
are designed to have a significant matter effect so that the hierarchy may be 
determined under favorable values of the unknown neutrino parameters. 
Although it is generally accepted that long-baseline experiments are 
the most promising method for 
determining the mass hierarchy, alternative ways should be explored
especially if they have the most sensitivity for neutrino parameters which
are complementary to the favorable parameters for long baseline experiments.

It was proposed in \cite{NPZ,GJK} that accurate measurement of
effective atmospheric scale $\Delta m^2$ both in electron and muon
neutrino disappearance channels would determine the neutrino mass
hierarchy, provided that precision of the measurement could reach sub
percent level. We call this the ``$\Delta m^2$ Disappearance'' method.
In this paper, we describe a setting by which such precision 
measurement could be achieved, and perform a semi-quantitative 
analysis to estimate the parameter region in which the mass 
hierarchy can be determined by this method.
For the measurement of $\Delta m^2$ in $\nu_{\mu}$ disappearance 
channel, which will be denoted as $\Delta m^2(\mu \mu)$, we 
assume a factor of 2-4 improvement with respect to what is currently 
expected for the phase II of the T2K project. 
For the measurement of $\Delta m^2(\text{ee})$, we assume an experiment 
based on the resonant absorption of $\bar{\nu}_{e}$ enhanced by 
the M\"ossbauer effect as recently proposed by Raghavan \cite{raghavan}.

After reviewing the  ``$\Delta m^2$ Disappearance'' method 
in Sec.~\ref{concept}, we discuss the accuracy of the 
determination of $\Delta m^2(\mu \mu)$ in Sec.~\ref{mumu}, and
recollect the one for $\Delta m^2(\text{ee})$ in Sec.~\ref{ee}.  In
Sec.~\ref{analysis} we present our analysis method and results.
Finally, in Sec.~\ref{conclusion} we draw our conclusions.

%%%%%%%%%%%%%%%%%%%%%%%%%%%%%%%%%%%%%%%%%%%%%%%%%%%%%%%%%%%%%%%%%%%%%%%%%%%
\section{Determining the neutrino mass hierarchy by $\nu_{e}$ and 
$\nu_{\mu}$ disappearance measurements - ``$\Delta m^2$ Disappearance'' method}
\label{concept}
%%%%%%%%%%%%%%%%%%%%%%%%%%%%%%%%%%%%%%%%%%%%%%%%%%%%%%%%%%%%%%%%%%%%%%%%%%%

Let us first explain the unconventional way of determining neutrino
mass hierarchy, the ``$\Delta m^2$ Disappearance'' method \cite{NPZ,GJK},
 to be explored in this paper.  Suppose
that we measure the atmospheric $\Delta m^2$ by doing disappearance
measurements of $\nu_{e} \rightarrow \nu_{e}/\bar \nu_{e} \rightarrow
\bar \nu_{e}$ and $\nu_{\mu} \rightarrow \nu_{\mu}/\bar\nu_{\mu}
\rightarrow \bar \nu_{\mu}$ in vacuum.
For clarity, we start with a simplified setting where the solar mixing
angle $\theta_{12}$ vanishes.  In this case, it is obvious that
$\nu_{e} \rightarrow \nu_{e}$ and $\nu_{\mu} \rightarrow \nu_{\mu}$
channels are governed by $\Delta m^2_{31}$ and approximately $ \Delta m^2_{32}$,
respectively, since the two oscillations scales approximately decouple
from each other.  Notice that because $m_{2} > m_{1}$, $|\Delta
m^2_{31}| > |\Delta m^2_{32}|$ for the normal hierarchy and $|\Delta
m^2_{31}| < |\Delta m^2_{32}|$ for the inverted hierarchy.  Therefore,
in the hypothetical simplified world of vanishing $\theta_{12}$, one
could, in principle, determine the mass hierarchy just by comparing
the absolute values of the two $\Delta m^2$, $|\Delta m^2_{31}|=\Delta
m^2(\text{ee})$ and $|\Delta m^2_{32}| \approx \Delta m^2(\mu \mu)$.

For non-zero $\theta_{12}$, the same considerations apply except that
the difference is reduced by approximately $\cos 2 \theta_{12}$.  To
make the discussion transparent we introduce, following \cite{NPZ},
the effective atmospheric $\Delta m^2(\alpha \alpha)$ determined by
the disappearance measurement in the $\nu_{\alpha} \rightarrow
\nu_{\alpha}$ channel. It is this effective $\Delta m^2 (\alpha
\alpha)$, obtained by a two-flavor approximate description of the full
three-flavor situation, which contains to high accuracy all the
three-flavor mixing effects.
Let us briefly review the derivation of $\Delta m^2(\alpha \alpha)$. 
In the approximate two-flavor mixing framework, 
the survival probability is given by 
\begin{eqnarray}
1-P(\nu_{\alpha} \rightarrow \nu_{\alpha}) = 
\sin^2 2\theta \; 
\sin^2 {\left(
\frac{\Delta m^2(\alpha \alpha) \,L}{4 \, E}
\right)}\, . 
\label{Palphaalpha2}
\end{eqnarray}
The energy $E_{\text{dip}}^{\text{2-flavor}}$ 
at which $1-P(\nu_{\alpha} \rightarrow
\nu_{\alpha})$ is maximum is given by $\displaystyle {\Delta
  m^2(\alpha \alpha) L}={2\pi E_{\text{dip}}^{\text{2-flavor}}} $.  Now we
repeat the same computation with the full three-flavor expression of
the survival probability; $E_{\text{dip}}^{\text{3-flavor}}$ can be
obtained by solving $d[1-P(\nu_{\alpha} \rightarrow
\nu_{\alpha}) ]/dE =0$.
By demanding that $E_{\text{dip}}^{\text{2-flavor}} =
E_{\text{dip}}^{\text{3-flavor}}$, an expression for $\Delta
m^2(\alpha \alpha)$ can be derived.  Assuming vacuum oscillations and
ignoring the terms of the order of $\sim (\Delta m^2_{21} / \Delta
m^2_{31})^2$, we obtain \cite{NPZ},
\begin{eqnarray}
\Delta m^2(\alpha \alpha) = r_{\alpha} |\Delta m^2_{31}| + 
(1-r_{\alpha}) |\Delta m^2_{32}|\, , 
\label{dm2eff}
\end{eqnarray}
where
\begin{eqnarray}
r_{\alpha} \equiv \frac{ |U_{\alpha 1}|^2 } { |U_{\alpha 1}|^2 + |U_{\alpha 2}|^2}
\, .
\label{defr}
\end{eqnarray}
The $U_{\alpha i}$'s are elements of the neutrino mixing matrix \cite{MNS},  
for which we use the standard parametrization \cite{PDG}.
The physical meaning of $r_\alpha$ is that it is the fraction of the
$\alpha$ flavor in the $\nu_1$ eigenstate over the sum of this fraction in 
the $\nu_1$ and $\nu_2$ eigenstates.  Note that $r_{e} = \cos^2\theta_{12}$
without further approximation and $r_{\mu} \approx \sin^2\theta_{12}$
if $\sin\theta_{13} \ll 1$.

The difference $\Delta_{\text{e}\mu}$ of the two effective 
$\Delta m^2$ can be easily computed from Eq.~(\ref{dm2eff}) and 
(\ref{defr}) to be
\begin{eqnarray}
\Delta_{\text{e}\mu} & \equiv & \Delta m^2 (\text{ee})  - \Delta m^2 (\mu \mu) 
 =  (r_e - r_\mu) (  |\Delta m^2_{31}| -|\Delta m^2_{32}|). 
\label{delta_emu}
\end{eqnarray}
Where 
\begin{eqnarray} 
r_e - r_\mu & = & \cos 2\theta_{12}
-\cos \delta \sin \theta_{13} \sin 2 \theta_{12} \tan \theta_{23} + {\cal O}(\sin^2 \theta_{13})\, , 
\end{eqnarray}
which is positive for small values of $\sin \theta_{13}$, 
and
\begin{eqnarray}
|\Delta m^2_{31}| -|\Delta m^2_{32}|  & = &  \pm \, \Delta m^2_{21}\, ,
\end{eqnarray}
where the $+ (-)$ sign is for the normal (inverted)
hierarchy.\footnote{SNO's demonstration \cite{SNO} that the solar CC/NC ratio is
  less than $\frac{1}{2}$ implies $\Delta m^2_{21}>0$.} 
  Thus the {\it sign} of $\Delta_{\text{e} \mu}$ is the same as the {\it sign} of
$\Delta m^2_{31}$.  Therefore, within the experimentally allowed
region for the mixing parameters \footnote{Except for some extreme
  values.}, the hierarchy is normal if $\Delta_{\text{e}\mu}$ is
positive and inverted if $\Delta_{\text{e}\mu}$ is negative.

One should be reminded that {\it both} measurements, $\Delta
m^2(\text{ee})$ and $\Delta m^2(\mu \mu)$, have to be accurate to the
1\% level at least, because $|\Delta_{\text{e} \mu}/\Delta m^2(\alpha
\alpha)|\sim$ 1\% for the current best fitted parameters.  A more
elaborate treatment in \cite{NPZ} entailed a rough estimation that
both accuracies must be better than 0.5\%.  Therefore, the issue is
whether one can at least envisage experimental setups capable of
measuring both $\Delta m^2$ to such a high accuracy.

%%%%%%%%%%%%%%%%%%%%%%%%%%%%%%%%%%%%%%%%%%%%%%%%%%%%%%%%%%%%%%%%%%%%%%%%%%%
\section{Precision measurement of $\Delta m^2(\mu \mu)$ by 
T2K II experiment}
\label{mumu}
%%%%%%%%%%%%%%%%%%%%%%%%%%%%%%%%%%%%%%%%%%%%%%%%%%%%%%%%%%%%%%%%%%%%%%%%%%%

It appears that precision determination of $\Delta m^2(\mu \mu)$ is
more feasible because there exists a well known method: the spectrum
measurement in the $\nu_{\mu}$ disappearance channel.  If it can be
performed with high precision, it should be possible to determine 
$\Delta m^2(\mu \mu)$ with great accuracy.

Unfortunately, there is a potential obstacle to accurate measurement
of $\Delta m^2(\mu \mu)$, the problem of {\em absolute energy scale
  uncertainty}.  For concreteness, we consider the phase II of the T2K
project (T2K II)~\cite{JPARC} where the beam power from J-PARC is
upgraded to 4 MW and the far detector will be Hyper-Kamiokande with
0.54 Mt fiducial volume.
%We assume the 2.5 degree off axis $\nu_\mu$ beam, which has a peak at around 0.65 GeV, and 2 years running in the neutrino mode only.
%
With huge number of events obtained in several years of running, 
the precision of $\Delta m^2(\mu \mu)$ would reach to 
$\sim 0.1$\% level without the energy scale uncertainty \cite{resolve23}. 
To determine $\Delta m^2$ accurately, however, we have to know
precisely the absolute value of the energy of the detected muons 
to reconstruct the neutrino energies. 
Since the oscillation probability depends on $\Delta m^2/E$, 
the accuracy of $\Delta m^2(\mu \mu)$ measurement is limited by 
the energy scale error. 
At the moment, the absolute energy of muons are known to the accuracy
of 2\% at energies $\sim 1$ GeV in the Super-Kamiokande
detector~\cite{SKI_atm}.
Lacking better calibration sources, one expects that the energy scale 
uncertainty will be the major limiting factor in improving the 
$\Delta m^2(\mu \mu)$ determination in the T2K II experiment \cite{private}. 
Notice that the difficulty is not specific to this experiment and 
similar statements would apply to other detection methods as well. 

%If this is the case, our method for determining the mass hierarchy

If the 2\% energy scale uncertainty still remains, our method for 
determining the mass hierarchy would not work because then 
the maximum precision we can expect on $\Delta m^2(\mu \mu)$ 
is 2\%. See Appendix \ref{bound}. 
A precision better than  1\% is needed for the 
method discussed in this paper to be useful.
In this work, we take the optimistic attitude that this uncertainty
can be reduced to the level of 0.5\%-1\% by the time T2K II will be
realized, or by development of alternative detection technologies.  We
study to what extent one can determine the mass hierarchy with the
method described in the previous section.  We believe that
%despite that it is not clear if this uncertainty can be reduced to the 
%required level,
physics results we can achieve under such an assumption are worthwhile
to report. 

For definiteness, we assume that $\Delta m^2(\mu \mu)$ can
be determined with 0.5\% accuracy. 
We also examine the case in which the accuracy  is 1\% to reveal 
how the capability of the mass hierarchy determination 
depends upon the accuracy of  $\Delta m^2(\mu \mu)$. 
We note that due to the fact that a large oscillation
effect in the $\nu_\mu \to \nu_\mu$ mode is expected, the precision in the 
determination of $\Delta m^2(\mu \mu)$ essentially 
does not depend on the precise values of the unknown mixing parameters 
such as $\theta_{13}$ and $\delta$, and possible deviation of $\theta_{23}$ 
from $\pi/4$. Therefore, the fractional accuracy, 
$\delta (\Delta m^2(\mu \mu))/\Delta m^2(\mu \mu)$ can be regarded, 
to good approximation, as constant for a given systematic uncertainty.
Also, since matter effects change the size of $\Delta m^2(\mu \mu)$ 
by less than 0.1\% they can be ignored for the T2K measurement of
$\Delta m^2(\mu \mu)$, see \cite{NPZ}.

%%%%%%%%%%%%%%%%%%%%%%%%%%%%%%%%%%%%%%%%%%%%%%%%%%%%%%%%%%%%%%%%%%%%%%%%%%%
\section{Precision measurement of $\Delta m^2(\text{ee})$ by  
M\"ossbauer enhanced resonant absorption of 
$\bar{\nu}_e$ }
\label{ee}
%%%%%%%%%%%%%%%%%%%%%%%%%%%%%%%%%%%%%%%%%%%%%%%%%%%%%%%%%%%%%%%%%%%%%%%%%%%

Most probably, the most difficult part of the method for the determination 
of the neutrino mass hierarchy explored in this paper is to measure 
$\Delta m^2(\text{ee})$ to the level of a few parts per mil. 
The reactor $\theta_{13}$ experiments \cite{reactor}, as they stand, 
will not reach the accuracy of 1\% level. 
Recently, the intriguing possibility of using 
the resonant absorption reaction \cite{mikaelyan}
\begin{eqnarray}
\bar{\nu}_{e} \, + \, ^{3}\mbox{He} + \mbox{orbital e}^{-} \, \rightarrow \; ^{3}\mbox{H},
\label{res-abs}
\end{eqnarray}
to explore $\bar{\nu}_{e}$ interaction below the threshold of the
charged-current absorption reaction $\bar{\nu}_{e} + p \rightarrow n +
e^{+}$, has been proposed by Raghavan \cite{raghavan}.  See
Ref.~\cite{visscher,schiffer} for earlier suggestions.
The resonance condition is automatically satisfied if the
$\bar{\nu}_{e}$ beam is prepared with the use of the inverse reaction
$^{3}\mbox{H} \rightarrow \, \bar{\nu}_{e}+\; ^{3}\mbox{He} +
\mbox{orbital e}^{-}$.  If both, the source and the target, atoms are
placed in a metal lattice and can enjoy the same environment the resonant
cross section can be enhanced by the M\"ossbauer effect by a factor of
$\sim 10^{11}$ \cite{raghavan}.
See also~\cite{Potzel_snow2006} where some potential difficulties 
of this experiment are discussed. 

Because the energy of $\bar{\nu}_{e}$ from the bound state beta decay
is so low, $E = 18.6$ keV, the first oscillation maximum (minimum of
$P(\bar{\nu}_{e} \rightarrow \bar{\nu}_{e})$) is reached at the
baseline distance
$L_{\text{\text{OM}} } = 9.2 \times ( \vert \Delta m^2_{31}\vert/ 2.5
\times 10^{-3} \text{eV}^2)^{-1}$ m.  Then, one can envisage
$\theta_{13}$ experiments with $\sim$10 m baseline~\cite{raghavan}.
With the M\"ossbauer enhanced cross section of $\sigma_{\text{res}}
\simeq 5 \times 10^{-32} \text{ cm}^2$ the rate $R \equiv N_{T}
f_{\bar{\nu}_{e}} \sigma_{\text{res}}$, with $N_{T}$ being the number
of target atoms and $f_{\bar\nu_{e}}$ the neutrino flux, is given by
\begin{eqnarray}
R_{\mbox{enh}} = 1.2 \times 10^{6}
\left(  \frac{S \; M_{T}}{1 \, \mbox{MCi} \cdot \mbox{100~g} }  \right) 
\left(  \frac{L}{10~\mbox{m} }  \right)^{-2} 
\mbox{day}^{-1}, 
\label{enhancedR}
\end{eqnarray}
that is, a million events a day by using 100 g of $^3$He target mass $M_T$ 
and assuming a source strength $S$ of 1 MCi. 

It is natural to expect that the monochromatic nature of the
$\bar{\nu}_{e}$ beam as well as the extremely high statistics allow
one to measure $\Delta m^2(\text{ee})$ to a high accuracy, as
demonstrated by the recent analysis in \cite{mina-uchi}.
For the purpose of the present analysis, we use one of the particular 
settings discussed in that paper, the one called Run IIB, 
which is defined as follows:

\vspace{0.3cm}
\noindent 
Run IIB: it consists of measurements at 10 different detector
locations; $L_n=(2n+1)\, L_{\text{OM}}/5$ for $n=1,...,10$, 
so that the entire period, from 0 to 2$\pi$, is covered. At each
location an equal number of $10^{6}$ events is to be collected.

\vspace{0.3cm}

\noindent 
The authors of \cite{mina-uchi} argued that if the direct counting of
produced $^3\text{H}$ atoms works, a movable detector technique would
allow for a relative systematic uncertainty as low as 0.2\%, and if
not it may be of the order of $\simeq$1\%.  If the former uncertainty
is realized, a sensitivity to $\Delta m^2(\text{ee})$ of $\simeq
0.3~(\sin^2 2\theta_{13} / 0.1)^{-1}$\% at 1$\sigma$ CL is possible.
For the latter uncertainty, the sensitivity is worse by about a factor
of four \cite{mina-uchi}.  In this work, we consider this systematic
uncertainty to be 0.2\%, and also discuss the case where it is 1\%.

%%%%%%%%%%%%%%%%%%%%%%%%%%%%%%%%%%%%%%%%%%%%%%%%%%%%%%%%%%%%%%%%%%%%%%%%%%%
\section{Analysis method and results}
\label{analysis}
%%%%%%%%%%%%%%%%%%%%%%%%%%%%%%%%%%%%%%%%%%%%%%%%%%%%%%%%%%%%%%%%%%%%%%%%%%%

\begin{figure}[htbp]
%\vglue -3.0cm
%\begin{center}
\hglue -0.5cm
\includegraphics[width=1.02\textwidth]{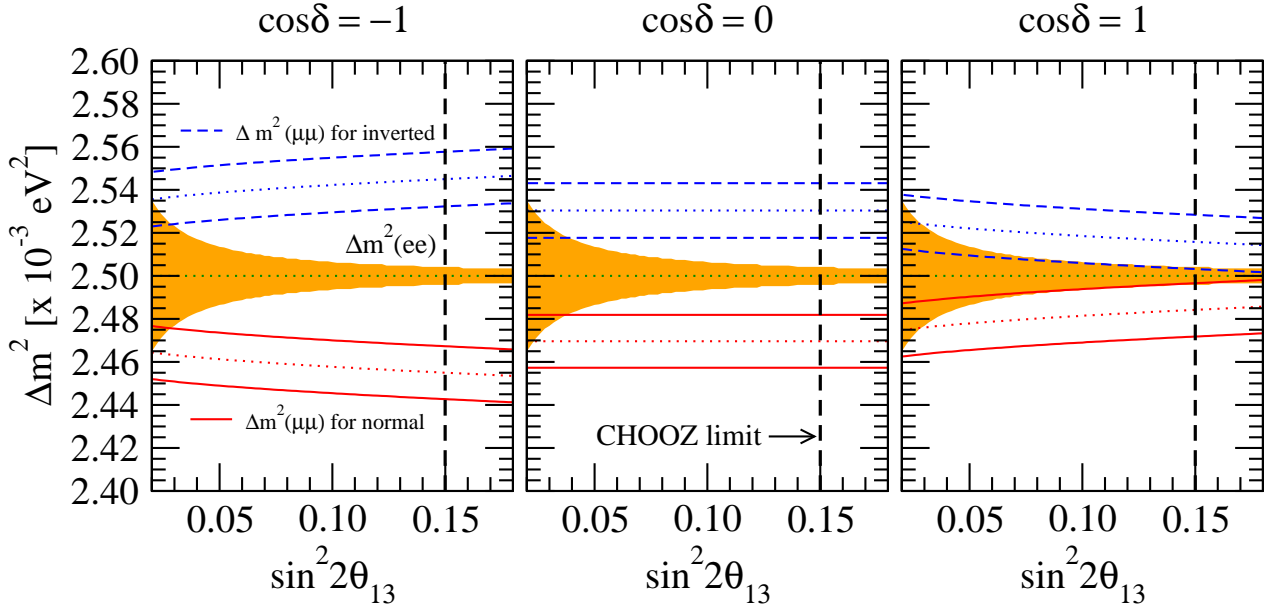}
%\end{center}
\vglue -0.4cm
\caption{Allowed regions for $\Delta m^2(\text{ee})$ (shaded by orange
  color) and $\Delta m^2(\mu \mu)$ (bands delimited by two solid and
  dashed curves) by measurement using the recoilless resonant
  $\bar{\nu}_{e}$ absorption reaction and the T2K II experiment,
  respectively, are plotted as functions of $\sin^2 2\theta_{13}$.
  The input value of $\Delta m^2(\text{ee})=2.5 \times 10^{-3}
  \text{eV}^2$ is assumed.  The red solid (blue dashed) curve 
for $\Delta m^2(\mu  \mu)$ denotes 
the case of normal (inverted) mass hierarchy.  
The left, the middle and the right panels are for 
the input values of 
$\delta=\pi$, $\delta=\pi/2$ or $3\pi/2$, 
and $\delta=0$ or $2\pi$,  respectively.  }
\label{1sigma-region}
\end{figure}

Given the accuracies of the measurements of $\Delta m^2(\mu \mu)$ and 
$\Delta m^2(\text{ee})$ discussed in Secs.~\ref{mumu} and \ref{ee}, 
it is straightforward to determine the region of mixing parameters 
in which the mass hierarchy can be resolved. 
To understand what are the relevant parameters,  we show in 
Fig.~\ref{1sigma-region}  the 1$\sigma$ CL determination of 
$\Delta m^2(\text{ee})$ and $\Delta m^2(\mu \mu)$ as a function 
of $\sin^2 2\theta_{13}$ for $\cos \delta = -1$ (left panel), 
0 (middle panel) and 1 (right panel). 
The physical parameters, $ \Delta m^2_{31}$ and $\Delta m^2_{32}$,  
are held fixed, which implies that $\Delta m^2(\text{ee})$ 
is also fixed. 
If the two strips of $\Delta m^2(\text{ee})$ and $\Delta m^2(\mu \mu)$ 
do not overlap then the mass hierarchy is determined at better than 84\% CL.
Even though they overlap the mass hierarchy can be determined 
to the extent that one can discriminate if 
$\Delta_{\text{e} \mu} \equiv \Delta m^2(\text{ee}) - \Delta m^2(\mu \mu)$
is positive (normal hierarchy) or negative (inverted hierarchy). 
Throughout this section we use the following values for  
the solar oscillation parameters:  
$\Delta m^2_{21}= 8.0\times 10^{-5}$eV$^2$ and 
$\sin^2 \theta_{12}=0.31$~\cite{SNO}, unless stated otherwise.

%%%%%%%%%%%% FIGURE I %%%%%%%%%%%%%%%
%%%%%%%%%%%% FIGURE I %%%%%%%%%%%%%%%

\vglue 0.2cm

A few remarks are in order: 

\vspace{0.2cm}
\noindent
(1) The dependence of the fractional uncertainty of $\Delta m^2(\text{ee})$ 
which is proportional to $(\sin^2 2\theta_{13})^{-1}$ \cite{mina-uchi} is 
clearly visible in Fig.~\ref{1sigma-region}. 

\vspace{0.2cm}
\noindent
(2) $\Delta m^2(\mu \mu)$ varies as a function of $\sin^2 2\theta_{13}$ 
because of the three-flavor effect in the disappearance probability 
$P(\nu_{\mu} \rightarrow \nu_{\mu})$, see Eq.~(\ref{delta_emu}).
Note, however, that the relative uncertainty with respect to its central 
value is independent of $\theta_{13}$. 

\vspace{0.2cm}
\noindent
(3) The three panels in Fig.~\ref{1sigma-region}, which correspond to 
different values of $\delta$, indicate that the discriminating sensitivity 
of the mass hierarchy depends upon $\delta$ in an interesting way. 
The sensitivity is highest (lowest) at $\delta = \pi$ (0 or $2\pi$), see  
Eq.~(\ref{delta_emu}).

To quantify the sensitivity region for the resolution of the mass
hierarchy we define the probability distribution function
$P_{\text{diff}} (\xi)$ of the difference $\xi \equiv \Delta m^2(\text{ee}) 
- \Delta m^2(\mu \mu)$.  Then the region of parameter 
which gives positive $\xi$ at $>$90, $>$95 and $>$99\% CL are determined 
by the condition
\begin{eqnarray}
\int_{0}^{\infty} d \xi \; P_{\text{diff}} (\xi) = 0.9, \; 0.95, \;  0.99 \, .
\label{def90}
\end{eqnarray}
Assuming that $\Delta m^2(\text{ee})$ and $\Delta m^2(\mu \mu)$ are  
Gaussian distributed\footnote{
%%%%%%%%%%%%%%%%% footnote %%%%%%%%%%%%%%%%%
  In good approximation, the $\chi^2$ distribution of $\Delta
  m^2(\text{ee})$ is Gaussian as far as we exploit the setting
  discussed in \cite{mina-uchi}.  }, $P_{\text{e}} (\Delta m^2 (\text{ee}))$ and 
$P_{\mu}(\Delta m^2(\mu \mu))$, with the average values 
$\overline{\Delta m^2(\text{ee})}$ and $\overline{\Delta m^2(\mu \mu)}$ and 
widths $\sigma_\text{e}$ and $\sigma_\mu$, respectively, $P_{\text{diff}}$ is also a
Gaussian distribution with average value $\overline{\Delta m^2(\text{ee})} - 
\overline{\Delta m^2(\mu \mu)}$ and 
width $\sqrt{\sigma_\text{e}^2 + \sigma_\mu^2}$.

\begin{figure}[htbp]
\vglue -0.50cm
\begin{center}
\hglue  -0.5cm
\includegraphics[width=0.55\textwidth]{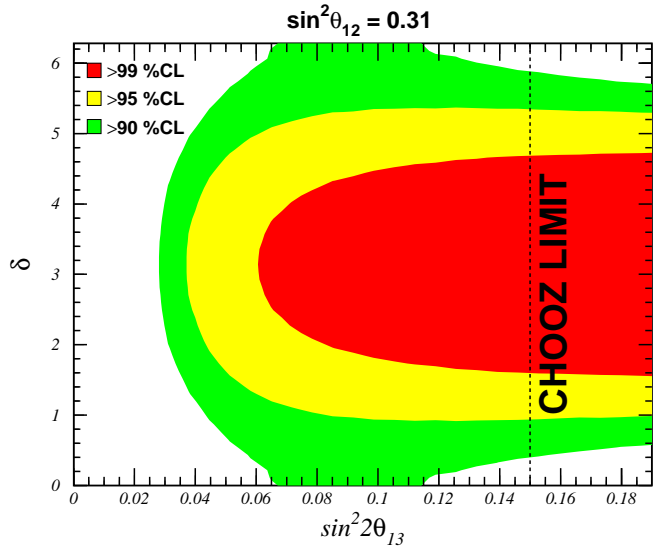}
\end{center}
\vglue -2.0cm
\caption{
Sensitivity regions in the $\sin^2 2\theta_{13} - \delta$ plane 
in which the mass hierarchy can be resolved 
at $>$90\% (green),  $>$ 95\% (yellow), and  $>$99\% (red) CL by 
the method of comparing the two disappearance measurements. 
The uncertainty on  $\Delta m^2(\text{ee})$ is roughly given
by $(0.3/\sin^2 2\theta_{13})\%$ under the assumed 0.2\% systematic
error and the uncertainty on 
$\Delta m^2(\mu \mu)$ is assumed to be 0.5\%.
Here the current best fit values $\sin^2 \theta_{12}= 0.31$, is used.
}
\label{sensitivity-region}
\end{figure}

Using the precision for the determination of $\Delta m^2(\mu \mu)$ and 
$\Delta m^2(\text{ee})$ obtained in Secs.~\ref{mumu} and \ref{ee}, 
it is straightforward to determine the sensitivity regions. 
In Fig.~\ref{sensitivity-region} we present the sensitivity regions 
in the space spanned by $\sin^2 2\theta_{13}$ and $\delta$
at $>$ 90\% (green),  $>$ 95\% (yellow), and  $>$ 99\% (red) CL in which 
the mass hierarchy can be resolved by the method of 
comparing these two disappearance measurements. 

%%%%%%%%%%%% FIGURE II %%%%%%%%%%%%%%%
%%%%%%%%%%%% FIGURE II %%%%%%%%%%%%%%%

As anticipated in the remark after Fig.~\ref{1sigma-region}, the
sensitivity depends significantly on the CP violating phase $\delta$.
It is highest at $\delta=\pi$, and lowest at $\delta=0$ or $2\pi$.  In
fact, this behavior is easy to understand from Eq.~(\ref{delta_emu}),
given that $\cos 2\theta_{12}$ is positive definite, the highest
(lowest) sensitivity is reached at $\cos \delta = -1$ (+1).

%%%%%%%%%%%% FIGURE III %%%%%%%%%%%%%%%
\begin{figure}[htbp]
\vglue -0.50cm
\begin{center}
\hglue  -0.5cm
\includegraphics[width=0.5\textwidth]{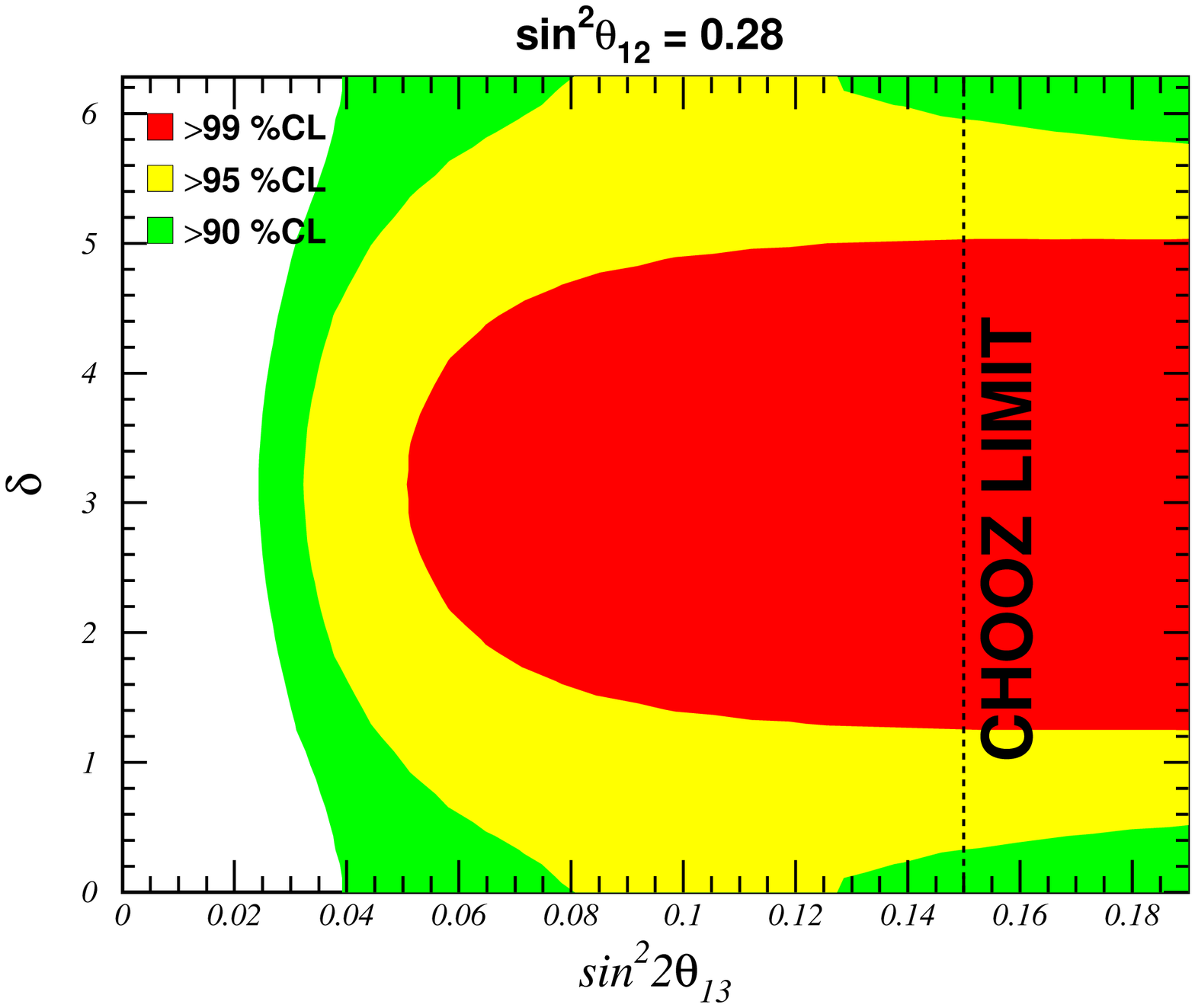}
\includegraphics[width=0.5\textwidth]{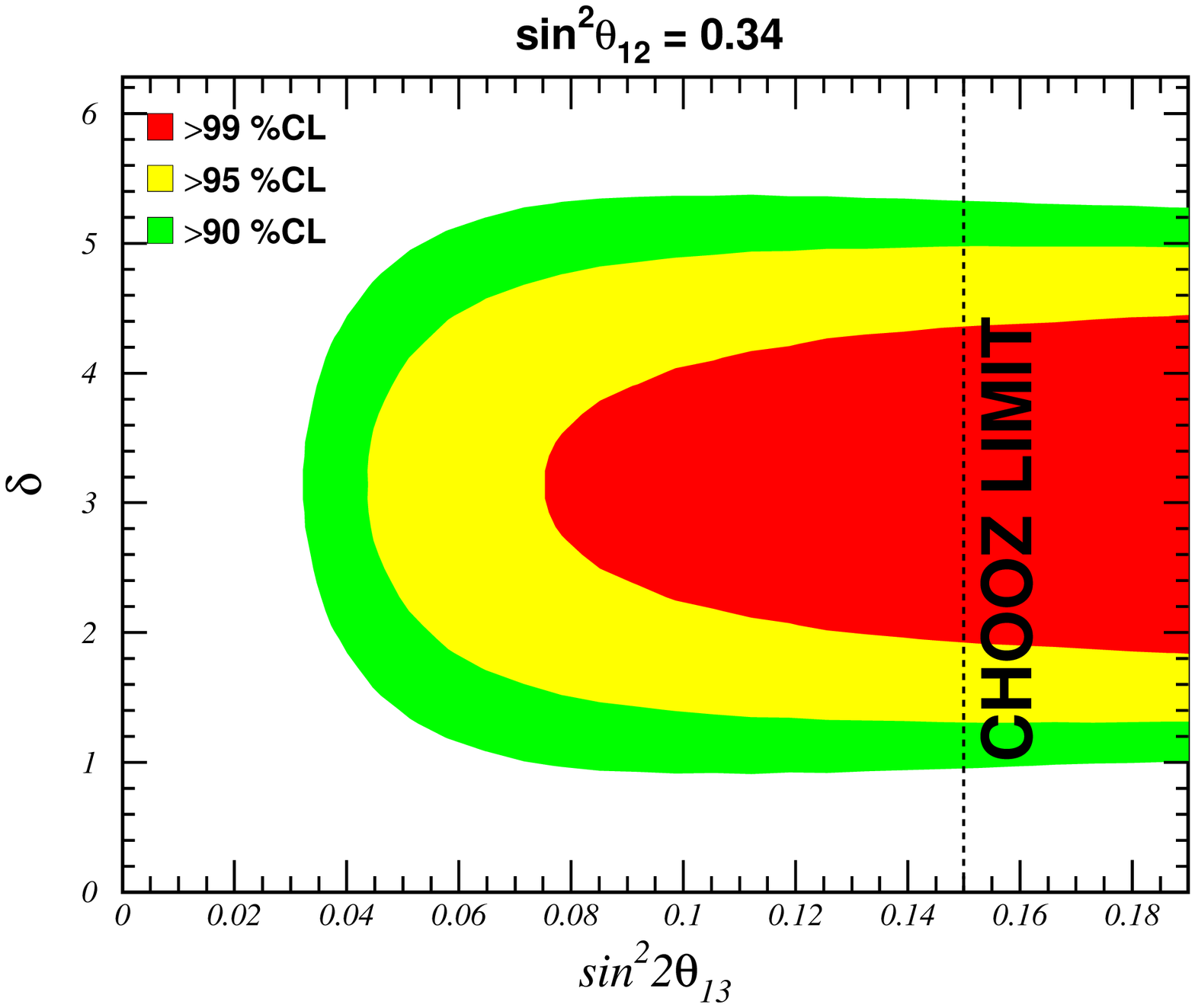}
\end{center}
\vglue -1.8cm
\caption{
Same as in Fig.~\ref{sensitivity-region} but for 
$\sin^2 \theta_{12}= 0.28$ (left panel) and 0.34 (right panel) 
which are allowed by the solar neutrino data and KamLAND at 2$\sigma$ CL. 
}
\label{sensitivity-region2}
\end{figure}
%%%%%%%%%%%% FIGURE III %%%%%%%%%%%%%%%

The strong dependence of the sensitivity on $\delta$ suggests that 
similar significant dependences exist also on other quantities, 
$\theta_{12}$, $\Delta m^2_{21} / \Delta m^2(\alpha \alpha)$ and 
$\tan \theta_{23}$. 
In Fig.~\ref{sensitivity-region2}, the sensitivity regions for determining 
the mass hierarchy are plotted for two values of $\theta_{12}$ different
from the best fit one but allowed by the current data at 2$\sigma$ CL.
It is remarkable that the sensitivity regions depend so strongly on 
$\theta_{12}$ which corresponds to
 $\cos 2\theta_{12}= 0.38 ^{+0.061}_{-0.072}$. 
The smaller $\theta_{12}$ the better the sensitivity. 
At small $\theta_{13}$ the second term in the right-hand side of
Eq.~(\ref{delta_emu}) is small, and the difference between the two
$\Delta m^2$ is governed by the first term.  The term is larger for
smaller $\theta_{12}$ in the first octant, and hence the sensitivity
is higher.
Similar argument applies to $\Delta m^2_{21} / \Delta m^2(\alpha
\alpha)$. If the true value of this ratio turns out to be larger
(smaller) than the current best fit, it will be easier (more
difficult) to determine the hierarchy.

We have presented the dependence of the sensitivity on $\theta_{12}$
because the significant improvement of the accuracy is not expected
for this quantity, unless either precision measurement of low-energy
solar neutrino flux is realized \cite{nakahata}, or a dedicated reactor
$\theta_{12}$ measurement is carried out \cite{sado}.
On the contrary, the accuracy of $\Delta m^2_{21}$ determination 
will be improved to about 5\% level by KamLAND alone   
as 10 times more statistics is accumulated by this experiment 
together with a better control of the systematic uncertainties~\cite{suzuki}. 
Of course, there is no need to discuss the dependence on 
$\Delta m^2 (\text{ee})$ and $\Delta m^2 ({\rm \mu \mu})$ because 
they must be determined with high accuracies to realize the method. 
To determine $\tan\theta_{23}$ accurately we need to solve 
the $\theta_{23}$ octant degeneracy by 
either the atmospheric \cite{atm23} or 
the reactor-accelerator combined methods \cite{resolve23}. 

%%%%%%%%%%%% FIGURE IV %%%%%%%%%%%%%%%
\begin{figure}[htbp]
\vglue -0.50cm
\begin{center}
\hglue  -0.5cm
\includegraphics[width=0.5\textwidth]{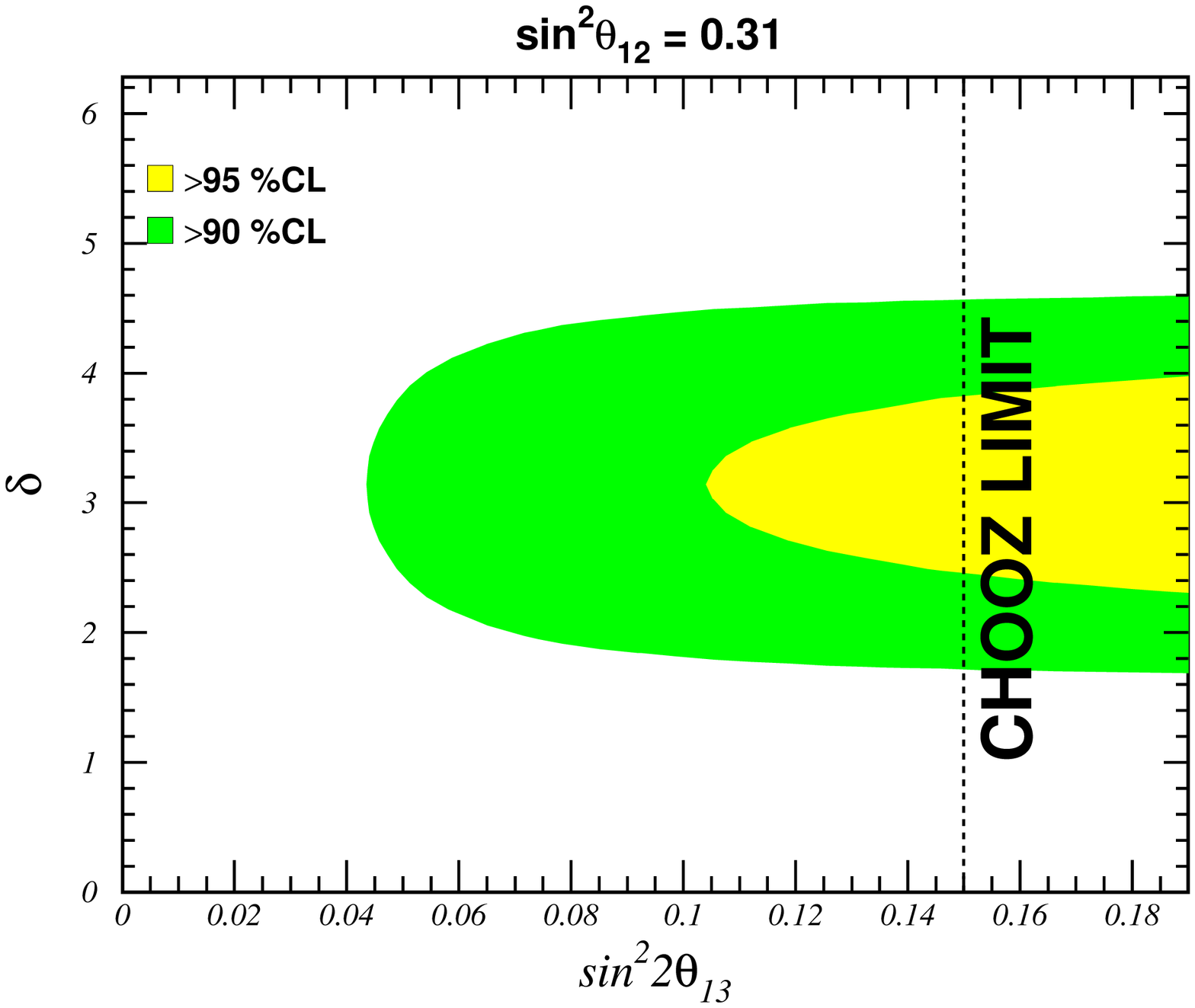}
\includegraphics[width=0.5\textwidth]{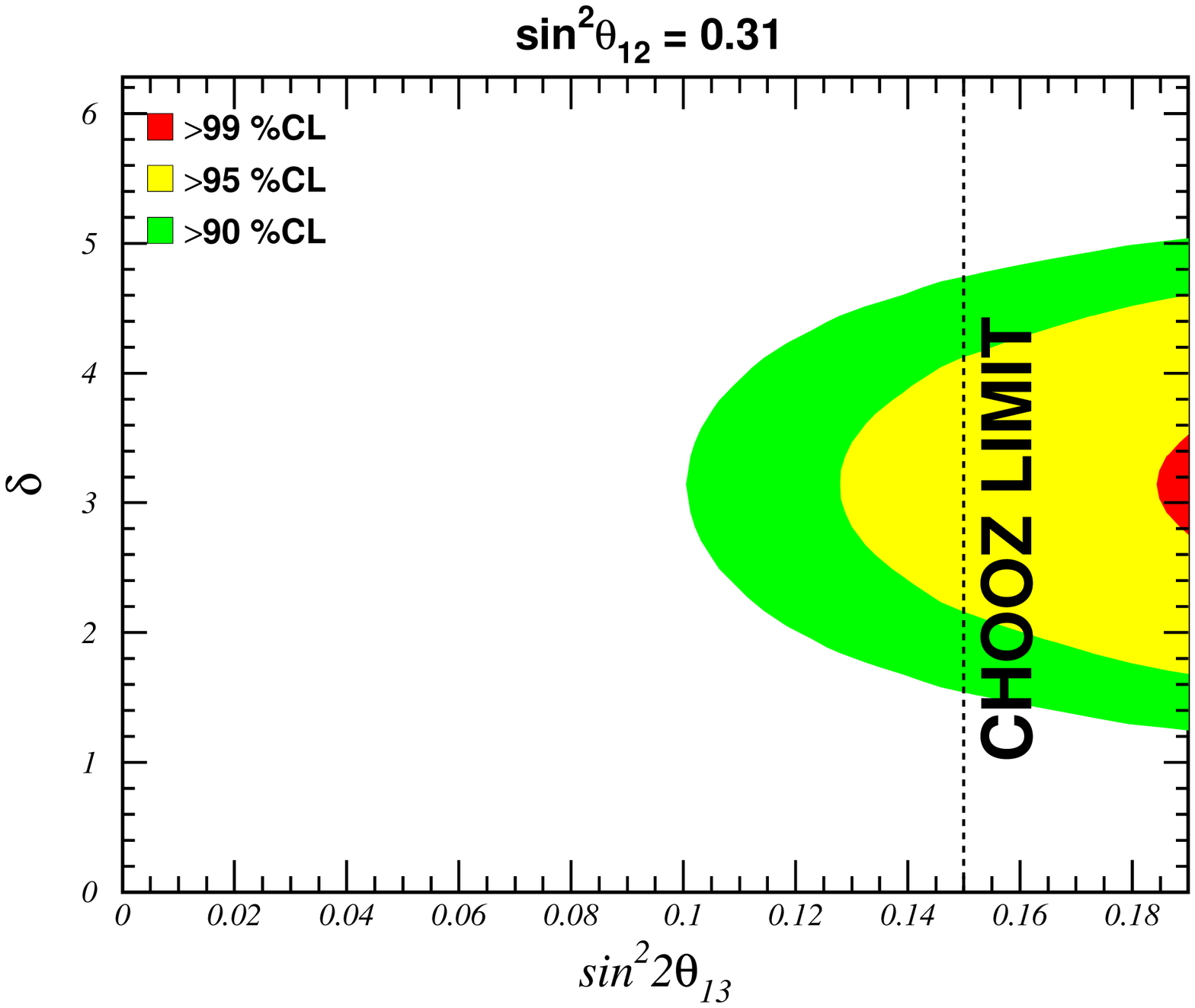}
\end{center}
\vglue -1.8cm
\caption{ Same as in Fig.~\ref{sensitivity-region} except: 
(left panel) the uncertainty on $\Delta m^2(\mu \mu)$ is increased by a
  factor of 2 to $1\%$, (right panel) sytematic error for the 
$\Delta  m^2(\text{ee})$ determination was increased by a factor of 5 
(from 0.2\% to 1\%) which implies that the uncertainty on 
$\Delta  m^2(\text{ee})$ is roughly described by 
$(1.2/\sin^2 2\theta_{13})\% $ with 
the uncertainty on $\Delta  m^2(\mu \mu)$ returned to $0.5\%$.  }
\label{sensitivity-region3}
\end{figure}
%%%%%%%%%%%% FIGURE IV %%%%%%%%%%%%%%%

Let us make some comments about the possibility to
determine the CP phase $\delta$.  Unfortunately, it
will not be possible to establish CP violation
($\sin \delta \neq 0$) by this method
without even further improvement on the precision with which 
both $\Delta m^2$'s are measured.
This can be understood by looking at Fig.~\ref{1sigma-region}.  
We can see that any point within the allowed region 
of $\Delta m^2(\mu \mu)$ delimited by the red solid (blue dashed) 
lines for $\cos \delta = 0 $ (corresponding to $\delta=\pi/2$ or 
$3\pi/2$, i.e., CP violation) can also be within the allowed 
region either for $\cos \delta = 1$ or $\cos \delta = -1$ 
(corresponding to no CP violation).  
However, if CP is not violated, there is some possibility to 
distinguish  $\delta =0$ from  $\delta=\pi$ since the two allowed 
regions for $\cos \delta = 1$ and $\cos \delta = -1$ have little 
overlap for $\sin^2 2\theta_{13} \gsim 0.05$.  It is
interesting to note that accelerator experiments like T2K II which
will operate at the first oscillation maximum will have difficulty to
distinguish the two points $\delta =0$ and $\pi$~\cite{KMN}.

Finally, let us also discuss the case where one of the assumption on
the precision for $\Delta m^2$ is not valid.  In the left panel of
Fig.~\ref{sensitivity-region3}, we show the same information as in
Fig.~\ref{sensitivity-region} but assuming that the precision for
$\Delta m^2(\mu \mu)$ is 1\% (instead of 0.5\%) but keeping the
accuracy for $\Delta m^2(\text{ee})$ the same as before. In this case
the sensitivity is significantly reduced compared to the case presented
in Fig.~\ref{sensitivity-region}, but there is still significant
parameter space where the mass hierarchy can be determined at $>$ 90\% CL.
On the other hand, in the right panel of
Fig.~\ref{sensitivity-region3}, we show the same information as in
Fig.~\ref{sensitivity-region} but assuming the systematic uncertainty for
the M\"ossbauer experiment is 1\% (instead of 0.2\%), which implies 
about the factor four larger uncertainty 
$\sim (1.2/\sin^2 2\theta_{13})\% $,  on $\Delta m^2 (\text{ee})$, 
but keeping the accuracy for $\Delta m^2(\mu \mu)$ the same as before
(0.5\%).  In this case,
there is much less parameter space where we can determine the mass
hierarchy.

%%%%%%%%%%%%%%%%%%%%%%%%%%%%%%%%%%%%%%%%%%%%%%%%%%%%%%%%%%%%%%%%%%%%%%%%%%%
\section{Concluding remarks}
\label{conclusion}
%%%%%%%%%%%%%%%%%%%%%%%%%%%%%%%%%%%%%%%%%%%%%%%%%%%%%%%%%%%%%%%%%%%%%%%%%%%

We have shown that if $\Delta m^2(\mu \mu)$ and $\Delta
m^2(\text{ee})$ can be measured, respectively, with 0.5\% and
$\sim 0.3~(\sin^2 2\theta_{13} / 0.1)^{-1}$~\% accuracy, the neutrino mass
hierarchy can be resolved down to $\sin^2 2\theta_{13} \sim 0.05$ for
$0.3 \, \pi \lsim \delta \lsim 1.7 \, \pi$ at $>$95\% CL for the current
best fit values of the oscillation parameters.  For slightly different
values of the oscillation parameters, still allowed by data, the
sensitivity can be even higher.  Although it is not possible to
determine the CP phase $\delta$ using these two measurements, it may
be possible to distinguish the case $\delta=0$ from $\delta=\pi$.

It should be stressed that high precision measurements of the muon and
electron disappearance oscillation experiments considered in this
paper have their own good physics motivation.  If these two
experiments will be realized, then as a special bonus, there is a
possibility that the mass hierarchy can be determined, even though 
neither of these experiments alone can accomplish this task.

Existing technologies seem to allow us to realize the experiment such
as the second phase of the T2K project, in which the determination of
$\Delta m^2(\mu \mu)$ up to 2\% precision looks quite feasible.  How
much one will be able to improve this precision depends essentially on
how much one can reduce the energy scale uncertainty.  
If it can be reduced to less than 1\%, then it is worthwhile to consider the
method studied in this paper.

The method discussed in this paper paper,
``$\Delta m^2$-Disappearance'' method, is most sensitive when
$\delta$ is near $\pi$. 
Whereas long baseline experiments 
%such as NO$\nu$A  
are most sensitive when $\delta=3\pi/2$ ($\pi/2$)
for the normal (inverted) mass hierarchy, \cite{MN01,MP}.  
%\cite{NOVA, MP}.  
Thus, these two methods are complimentary in the CP violating phase
variable, $\delta$.

On the other hand, it is not yet clear if the oscillation experiment
based on the M\"ossbauer enhanced resonant $\bar{\nu}_{e}$ absorption
can be really performed.  If it will be realized, the direct counting
of $^3$H atoms in the target is quite essential to carry out the
measurement of $\Delta m^2(\text{ee})$ to the required accuracies.

To measure both $\Delta m^2(\text{ee})$ and $\Delta m^2(\mu \mu)$ 
to the accuracy less than 1\% is challenging, but if it is 
possible, this will provide a direct and alternative way to determine
the neutrino mass hierarchy.

%%%%%%%%%%%%%%%%%%%%%%%%%%%%%%%%%%%%%%%%%%%%%%%%%%%%%%%%%%%%%%%%%%%%%%%%%%%
%%%%%%%%%%%%%%%% acknowledgments %%%%%%%%%%%%%
\begin{acknowledgments}
H.M. is supported by Bilateral Programs, Scientist Exchanges based on 
Japan Society of Promotion of Science and Brazilian Academy of Sciences.
He is grateful to Departamento de F\'{\i}sica, 
Pontif{\'\i}cia Universidade Cat{\'o}lica (PUC) do Rio de Janeiro 
for hospitality where this work was initiated and 
largely completed during his visit. 
We thank E.~Lisi, T.~Kajita, T.~Nakaya, W.~Potzel and 
T.~Schwetz for useful discussions and correspondences. 
This work was supported in part by the Grant-in-Aid for Scientific
Research, No. 16340078, Japan Society for the Promotion of Science,
Funda\c{c}\~ao de Amparo \`a Pesquisa do Estado de S\~ao Paulo
(FAPESP), Funda\c{c}\~ao de Amparo \`a Pesquisa do Estado de Rio de
Janeiro (FAPERJ) and Conselho Nacional  de Ci\^encia e
Tecnologia (CNPq). Fermilab is operated under DOE Contract No. 
DE-AC02-76CH03000.

\end{acknowledgments}
%%%%%%%%%%%%%%%%%%%%%%%%%%%%%%%%%%%%%%%%%%%%%%%%%%%%%%%%%%%%%%%%%%%%%%%%%%%

\newpage

\appendix

\section{The uncertainty on  $\Delta m^2$ is bounded from below 
by the absolute energy scale uncertainty}
\label{bound}

We assume, as an idealistic situation, that the neutrino beam is
monitored by a near detector with identical detection apparatus to a
far detector for measuring modulation of the neutrino spectrum.  Also,
the neutrino source is assumed to be a point source for both the near
and far detectors.  With the energy scale uncertainty $\Delta E$, the
true and the reconstructed neutrino energies, $E_{\text{true}}$ and
$E_{\text{rec}}$, are related in both detectors by $E_{\text{rec}} =
E_{\text{true}} + \Delta E = (1+x) E_{\text{true}}$ where $x \equiv
\Delta E/ E_{\text{true}}$.
For simplicity, we ignore all the other uncertainties except for the 
energy scale one and assume that the fractional uncertainty $x$ is 
constant over the energy range of interest. 
In this case, it is clear that the number of events observed 
per unit of detector mass at the near and far detectors, 
$N_\text{far}$ and $N_\text{near}$, respectively, are related to the 
neutrino disapperance probability $P$ by

\begin{eqnarray}
N_{\text{far}}(E_\text{rec}) &=&
\left(\frac{L_{\text{near}}}{L_\text{far}}\right)^2 \, 
N_{\text{near}}(E_\text{rec})\,
P \left(
\frac{\Delta m^2_{\text{true}} L_{\text{far}}}{E_{\text{true}}} 
\right)  
\nonumber \\
&=&
\left(\frac{L_{\text{near}}}{L_\text{far}}\right)^2 \, 
N_{\text{near}}(E_\text{rec})\,
P \left(
\frac{\Delta m^2_{\text{fit}} L_{\text{far}}}{E_{\text{rec}}} 
\right)
\, ,
\label{Nfar}
\end{eqnarray}
where $L_{\text{near}}$ ($L_{\text{far}}$) is the near (far) baseline.
This  implies

\begin{eqnarray}
\Delta m^2_{\text{fit}} = \left(\frac{E_{\text{rec}}}{E_\text{true}}\right)
\Delta m^2_{\text{true}} = (1+x) \Delta m^2_{\text{true}}\, .
\label{dmass}
\end{eqnarray}
So it is obvious that the percentage energy scale uncertainty sets a 
lower bound on the percentage uncertainty  of the $\Delta m^2$ determination.

%%%%%%%%%%%%%%%%%%%%%%%%%%%%%%%%%%%%%%%%%%%%%%%%%%%%%%%%%%%%%%%%%%%%%%%%%%%
% Bibliography
%%%%%%%%%%%%%%%%%%%%%%%%%%%%%%%%%%%%%%%%%%%%%%%%%%%%%%%%%%%%%%%%%%%%%%%%%%%

%%%%%%%%%%%%%%%%%%%%%%%%%%%%%%%%%%%%%%%%%%%%%%%%%%%%%%%%%%%%%%%%%%%%%%%%%%%
\end{document}